\documentclass[aps,prl,showpacs,preprint,superscriptaddress]{revtex4}
\usepackage{bm,color}
\usepackage{graphicx}
\usepackage{amsmath}
\usepackage{color}
\usepackage{bm}
\usepackage{amsmath,amssymb}
\newcommand{\ex}{{\rm (ex)}}
\newcommand{\so}{{\rm (so)}}
\newcommand{\ext}{{\rm ext}}
\newcommand{\A}{{\mathcal A}}
\newcommand{\Av}{{\bm{\mathcal A}}}
\newcommand{\T}{{\mathcal T}}
\newcommand{\Tv}{{\bm{\mathcal T}}}

\newcommand{\js}{j_{\rm s}}
\newcommand{\jsv}{{\bm j}_{\rm s}}
\newcommand{\sv}{{\bm s}}
\newcommand{\rv}{{\bm r}}
\newcommand{\kv}{{\bm k}}

\newcommand{\qv}{{\bm q}}
\newcommand{\w}{\omega}
\newcommand{\W}{\Omega}
\newcommand{\Ek}{\varepsilon_{\bm k}}
\newcommand{\pv}{{\bf p}}
\newcommand{\pauli}{\hat{\sigma}}
\newcommand{\pauliv}{\hat{\bm \sigma}}
\newcommand{\del}{\partial}
\newcommand{\grad}{{\bm \nabla}}

\newcommand{\me}{m_{\rm e}}
\renewcommand{\a}{\alpha}

\newcommand{\D}{J_{\rm ex}}

\newcommand{\E}{\varepsilon}
\newcommand{\Ef}{\varepsilon_{\rm F}}
\newcommand{\kf}{k_{\rm F}}
\newcommand{\g}{\hat{g}}
\newcommand{\ga}{g^{\rm a}}
\newcommand{\gr}{g^{\rm r}}
\renewcommand{\Im}{{\rm Im}}
\renewcommand{\Re}{{\rm Re}}
\renewcommand{\H}{{\cal H}}
\newcommand{\Mv}{{\bm M}}

\newcommand{\nv}{{\bm n}}

\newcommand{\mv}{{\bm m}}

\newcommand{\zv}{\hat{\bm z}}
\newcommand{\U}{\hat{U}}
\newcommand{\R}{{\mathcal R}}
\newcommand{\tr}{{\rm Tr}}
\newcommand{\Hv}{{\bm H}}
\newcommand{\ag}{\alpha_{\rm G}}
\begin{document}
\title{Electrically driven spin torque and dynamical Dzyaloshinskii-Moriya interaction in magnetic bilayer systems}
\author{Akihito Takeuchi}
\affiliation{Department of Physics and Mathematics, Aoyama Gakuin University, Sagamihara, Kanagawa 229-8558, Japan}
\author{Shigeyasu Mizushima}
\affiliation{Department of Physics and Mathematics, Aoyama Gakuin University, Sagamihara, Kanagawa 229-8558, Japan}
\author{Masahito Mochizuki}
\affiliation{Department of Applied Physics, Waseda University, Okubo, Shinjuku-ku, Tokyo 169-8555, Japan}
\affiliation{PRESTO, Japan Science and Technology Agency, Kawaguchi, Saitama 332-0012, Japan}
\begin{abstract}
Efficient control of magnetism with electric means is a central issue of current spintronics research, which opens an opportunity to design integrated spintronic devices. However, recent well-studied methods are mostly based on electric-current injection, and they are inevitably accompanied by considerable energy losses through Joule heating. Here we theoretically propose a way to exert spin torques into magnetic bilayer systems by application of electric voltages through taking advantage of the Rashba spin-orbit interaction. The torques resemble the well-known electric-current-induced torques, providing similar controllability of magnetism but without Joule-heating energy losses. The torques also turn out to work as an interfacial Dzyaloshinskii-Moriya interaction which enables us to activate and create noncollinear magnetism like skyrmions by electric-voltage application. Our proposal offers an efficient technique to manipulate magnetizations in spintronics devices without Joule-heating energy losses. 
\end{abstract}
\maketitle

\section*{Introduction}
Recent research in spintronics~\cite{Ohno98,Wolf01,Zutic04,Chappert07,Manchon15,Endoh18,Jungwirth18,Baltz18} is seeking efficient ways to control magnetism in materials and realize high-performance magnetic devices beyond conventional techniques based on classical electromagnetism. To manipulate the magnetization in magnets, we need to exert torques to drive them. Spin-transfer torque is one such torque and, mediated by the exchange interaction, originates with the angular momentum transfer from the spins of the conducting electrons to magnetizations in the magnetic structure~\cite{Slonczewski96,Berger96,Tatara08}. Another is the so-called $\beta$-term torque originating from the spin relaxation due to the spin-orbit interactions and/or magnetic impurity scatterings, which is perpendicular to the spin transfer torque and thus behaves as a dissipative torque~\cite{Tatara08,Zhang04,Thiaville05}. These two torques enable the magnetization to be driven and switched via the injection of electric currents~\cite{Tsoi03,Klaui03,Yamaguchi04,Yamanouchi04}.

The Rashba-type spin-orbit interaction (RSOI) has recently attracted interest as a means to manipulate conduction-electron spins~\cite{Rashba60,Manchon15}. This interaction is of relativistic origin and becomes active in systems with broken spatial inversion symmetry such as surfaces and interfaces~\cite{Rashba60,Manchon15,Nitta97,Ast07,Nakagawa07}. Because the RSOI mediates mutual coupling between spins and orbital momenta of the conduction electrons, it works as an effective magnetic field acting on their spins. The strength and direction of the effective magnetic field that each electron feels are governed by the momentum of the electron. Therefore, the RSOI can be a source of nontrivial torques acting on the magnetization through its control on the spin polarizations of the conduction electrons~\cite{Obata08,Manchon08,Manchon09,Kim12,Miron10,Miron11}. The strength of the RSOI can be tuned by applying a gate voltage normal to the surface or interfacial plane~\cite{Nitta97}, which modulates the extent of the spatial inversion asymmetry. Thus, an alternate-current (AC) gate voltage produces a time-varying RSOI, and thereby offers a potential technique to produce electrically an oscillating magnetic texture via the nontrivial Rashba-mediated torques. This technique must be useful for manipulation of magnetic skyrmions in a magnetic bilayer system, which have recently attracted great interest from the viewpoint of possible application to high-performance memory devices~\cite{Fert13,Soumyanarayanan16}.

Recent theoretical studies demonstrated that this time-varying RSOI induces an AC spin current~\cite{Malshukov03,Zhang13,Ho14}. Ho {\it et al.} derived an expression for the torque $\Tv$ from the AC spin current~\cite{Ho15}, finding that this expression does not contain a term corresponding to the $\beta$-term torque. It is known that the spin-transfer torque alone cannot drive magnetic domain walls, but other types of torques such as the $\beta$-term torque and the spin-orbit torque~\cite{Ando08, Emori13, RyuKS13} are required to drive them. However, their theory was based on a continuity equation for the conduction electron spins where the contribution of spin relaxation torque is totally neglected. The lack of a $\beta$-term torque may be a consequence of this crude approximation~\cite{Tatara08PRB}. Therefore, we reexamine whether the $\beta$-term torque in the time-dependent Rashba electron systems is present using a more elaborate theoretical method.

In this paper, we derive using the quantum field theory an explicit analytical formula for the torque arising from the AC RSOI in the magnetic bilayer systems. We demonstrate that the appropriate incorporation of the spin relaxation effect leads to a formula including both the spin-transfer torque and the $\beta$-term  torque given in terms of the spin current proportional to the time derivative of the RSOI. We find that the derived expression may be regarded as a time-dependent Dzyaloshinskii-Moriya interaction (DMI) at the interface and may be exploited to manipulate noncollinear magnetic textures such as domain walls, magnetic vortices, and skyrmions. For a demonstration, we performed a numerical simulation of the electrical activation of a two-dimensional skyrmion lattice, where the effective AC DMI induces a periodic expansion and contraction of the skyrmions. This collective excitation of the skyrmions resembles the breathing mode~\cite{Mochizuki12,Onose12} observed in a skyrmion crystal under out-of-plane microwave magnetic fields. Efficient techniques to control and drive magnetism using electric fields are subject of intensive studies in the recent spintronics research because electric fields in insulators are not accompanied by Joule-heating losses. Note that electric currents and resulting Joule-heating losses can be significantly suppressed by using a ferromagnet/insulator bilayers or a metallic bilayer fabricated on an insulating substrate. Our finding provides a new efficient technique to control magnetism using AC electric fields and means to electrically excite eigenmodes of noncollinear spin textures.

\section*{Results}
\subsection*{Model}
\begin{figure}[tb]
\centering
\includegraphics[scale=1.0, clip]{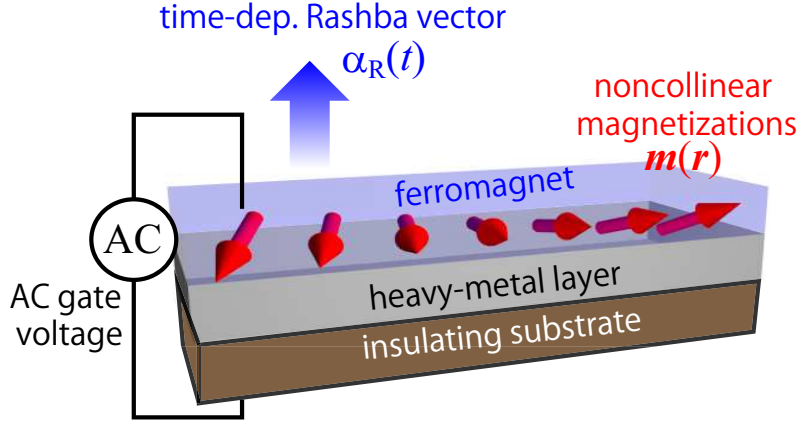}
\caption{Schematic illustration of the time-dependent Rashba electron system interacting with local magnetizations $\mv(\rv)$. The Rashba parameter $\a_{\rm R}(t)$ is time-modulated by an external AC electric voltage. The insulating substrate prevents electric-current flows and enhances the effects of the electric voltage acting on the RSOI-hosting interface.}
\label{fig1}
\end{figure}

We consider a magnetic bilayer system, that is, a two-dimensional electron system on top of the surface of a magnet (see Fig.~\ref{fig1}). This system is fabricated on an insulating substrate to enhance the effects of the electric gate voltage acting on the ferromagnet/heavy-metal interface through preventing the electric-current flow. The total Hamiltonian of this electron system has four contributing terms, $H = H_{\rm K} +H_{\rm R} +H_{\rm ex} +H_{\rm imp}$ with
\begin{align}
H_{\rm K}
&=
\frac{1}{2 \me} \int{d^2r} \, \big| \pv \psi(\rv,t) \big|^2
-\Ef \int{d^2r} \, \psi^\dagger(\rv,t) \psi(\rv,t),
\\
H_{\rm R}
&=
-\frac{\a_{\rm R}(t)}{\hbar} \int{d^2r} \, \psi^\dagger(\rv,t) (\pv \times \pauliv)_z \psi(\rv,t),
\\
H_{\rm ex}
&=
\D \int{d^2r} \, \mv(\rv) \cdot \psi^\dagger(\rv,t) \pauliv \psi(\rv,t),
\\
H_{\rm imp}
&=
\int{d^2r} \, v_{\rm imp}(\rv) \psi^\dagger(\rv,t) \psi(\rv,t),
\end{align}
where $\me$ and $\pv$ denote the mass and momentum, respectively, of a conduction electron, $\Ef$ the Fermi energy, $\pauliv$ the Pauli matrices, and $\psi^{\dagger}$ ($\psi$) the creation (annihilation) operator of a conduction electron. The term $H_{\rm K}$ represents the kinetic energies of the conduction electrons with $\varepsilon_{\rm F}$ being the Fermi energy, while the term $H_{\rm R}$ describes the time-varying RSOI with $\a_{\rm R}(t)$ being the time-dependent coupling coefficient. The term $H_{\rm ex}$ represents the exchange interaction between the conduction-electron spins and the local magnetization with $\D$ and $\mv$ being the coupling constant and the normalized local magnetization vector, respectively. The term $H_{\rm imp}$ represents the scattering potentials from spatially distributed nonmagnetic impurities and determines the relaxation time of electrons given by $\tau$. More conretely, we consider the impurity potential given by $v_{\rm imp}(\rv) = u_{\rm imp} \sum_i \delta(\rv -{\bm R}_i)$, where $u_{\rm imp}$ denotes the strength of impurity scattering, ${\bm R}_i$ positions of the impurities, and $\delta(\rv)$ the Dirac delta function. When we take an average over the impurity positions as $\overline{v_{\rm imp}(\rv)} = 0$ and $\overline{v_{\rm imp}(\rv) v_{\rm imp}(\rv')} = n_{\rm imp} u_{\rm imp}^2 \delta(\rv-\rv')$, the relaxation time of electrons is given by $\tau = \hbar / 2 \pi \nu_{\rm e} n_{\rm imp} u_{\rm imp}^2$ in the first Born approximation. Here, $n_{\rm imp}$ denotes the concentration of impurities and $\nu_{\rm e}=\me/2\pi \hbar^2$ the density of state. In the present study, we focus on the spin torques induced by the time-varying RSOI but neglect those induced by the magnetization dynamics. Note that there should be a feedback effect that the spin states of the conduction electrons are modulated by the magnetization dynamics, but we neglect this subsequent effect on the spin torques.

\subsection*{Spin torque arising from time-dependent RSOI}
The torque induced by the electron spins via the exchange interaction is defined as
\begin{equation}
\Tv = \frac{\D a^2}{\hbar} \mv \times \sv,
\end{equation}
where $a$ is the lattice constant, $\sv=\langle{\psi^\dagger \pauliv \psi}\rangle$ is the conduction electron spin density, and the brackets denote the quantum expectation value. We assume a metallic bilayer system in which the condition $\D < \Ef$ usually holds and an adiabatic case with a slowly varying magnetization texture of $q \ll \kf$ where $q$ is the wavenumber of local magnetization and $\kf$ is the Fermi wavenumber. We also assume that a weak magnitude of $\a_{\rm R}$ ($\a_{\rm R} \kf \ll \Ef$) and low frequencies $\W$ ($\hbar \W \ll \Ef$) for the time-dependent RSOI. In this perturbation regime, we obtain the analytical formula of the torque in the form
\begin{equation}
\Tv=\Tv_1+\Tv_2+\Tv_3=
-\frac{a}{\hbar} D_1 (\mv \times \grad)_z \mv
+\frac{a}{\hbar} D_2(\mv \times \grad)_z \mv
-\frac{a}{\hbar} \beta_{\rm R} D_2 \mv \times \Big[ (\mv \times \grad)_z \mv \Big],
\label{eq:torque}
\end{equation}
where each coefficient is defined in the condition with $\Ef \gg \hbar / \tau$, $\D \gg \hbar / \tau$ and $\Ef -\D \gg \hbar / \tau$ as ($\eta = \hbar / 2 \tau$)
\begin{align}
D_1(t)
&=
\frac{\hbar \nu_{\rm e} a}{2 \pi \tau}
\bigg[ \frac{\Ef}{\D} \ln{\bigg(\frac{\Ef+\D}{\Ef-\D}\bigg)} -2 \bigg] \a_{\rm R}(t),
\label{eq:D1}
\\
D_2(t)
&=
\nu_{\rm e} a \Ef \tau
\frac{\D^2 (\D^2 -\eta^2)}{(\D^2+\eta^2)^2} \frac{d {\a}_{\rm R}(t)}{dt},
\label{eq:D2}
\\
\beta_{\rm R}
&=
\frac{2 \D \eta}{\D^2 -\eta^2}.
\label{eq:beta}
\end{align}
We note that $D_1$ vanishes in the clean limit with $\tau \to \infty$.
Note that $D_1$ does not vanish in the three-dimensional case or in the half-metallic case with $\D > \Ef$~\cite{Ado18} even in the clean limit.

In Eq.~\eqref{eq:torque}, the first two contributions, $\Tv_1+\Tv_2$, describe an effective DMI~\cite{Dzyaloshinskii57,Moriya60}, which is given in the continuum form as
\begin{equation}
\H_{\rm DMI}
=
\frac{D_1 -D_2}{a} \epsilon_{\alpha \beta z} \int{d^2 r} \, (\mv \times \nabla_\alpha \mv)^\beta.
\label{eqn:HDMI}
\end{equation}
(Note that $\mv \times (a^2/\hbar) \delta \H_{\rm DMI} / \delta \mv$ leads to $\Tv_1+\Tv_2$). The contribution $D_1$ ($\propto \a_{\rm R}$) appears even in the steady Rashba system~\cite{Ado18,Kim13,Kikuchi16}. Recent experiments indeed demonstrated voltage-induced variations of the interfacial DMI~\cite{Nawaoka15,Srivastava18}, which were ascribed to this steady Rashba contribution. Quite recently, an experimental observation of gigantic variation of the DMI that reaches 130~$\%$ has been reported for the Ta/FeCoB/TaO$_x$ multilayer system~\cite{Srivastava18}. In contrast, the contribution $D_2$ ($\propto \partial_t {\a}_{\rm R}$) appears only in a driven Rashba system with a time-dependent RSOI.

This interfacial DMI may be tuned by an electric gate voltage via the RSOI, and, more interestingly, an oscillating DMI may be achieved by applying an AC gate voltage. The Rashba coefficient $\a_{\rm R}(t)$ in the driven Rashba system is a sum of steady and time-dependent components, $\a_{\rm R}(t)=\a_0 + \a_\ext(t)$ with $\a_\ext(t) = \a_\ext \sin{(\W t)}$. For metallic (semiconducting) bilayer systems, the strength of this Rashba-induced DMI is roughly estimated to be
$D_1 \sim 0.1$~meV ($6 \times 10^{-6}$~meV) and
$D_2 \sim 5 \times 10^{-3}$~meV ($2 \times 10^{-6}$~meV).
Here we assume typical parameter values~\cite{Nitta97,Ast07,Nakagawa07}, i.e.,
$a=5$~{\AA},
$\Ef=4$~eV ($10$~meV),
$\kf=1$~\AA$^{-1}$ ($0.01$~\AA$^{-1}$),
$\D/\Ef=0.25$ ($0.5$),
$\tau=10^{-14}$~s ($10^{-12}$~s),
$\a_0=2$~eV$\cdot${\AA} ($0.07$~eV$\cdot$\AA),
$\a_\ext / \a_0 = 0.1$,
and $\W / 2 \pi = 1$~GHz.
The strength of the Rashba-mediated DMI is relatively strong (weak) in metal (semiconductor) systems. We also note that the magnitude of $D_2$ being proportional to $\partial_t \a_\ext(t)$ may be tuned by changing the amplitude and frequency of the AC gate voltage. The relative strength of $D_2$ or the ratio $D_2 / D_1$ tends to be small. Note that the ratio $D_2/D_1$ is approximately given by $\Omega \varepsilon_{\rm F} \tau^2/2\pi \hbar$, which takes $\sim 10^{-4}$ ($10^{-2}$) for metallic (semiconducting) bilayer systems when a typical frequency of $\Omega$=1 GHz is assumed. Namely, the ratio $D_2/D_1$ tends to be larger for the semiconducting system, whereas the absolute value of $D_2$ tends to be larger for the metallic system. We should choose an appropriate system depending on the target phenomena or the experiments.

In Eq.~\eqref{eq:torque}, the last two terms proportional to $D_2$ may be rewritten as
\begin{equation}
\Tv_2+\Tv_3
\propto
(\jsv \cdot \grad) \mv
-\beta_{\rm R} \mv \times (\jsv \cdot \grad) \mv.
\end{equation}
by adopting a definition of the spin current $\jsv \equiv (e / \hbar a) D_2 \zv \times \mv$. We find that these terms have equivalent forms with the spin-transfer torque and the nonadiabatic torque associated with the spin current $\jsv$, respectively.
With an AC-dependent $\a_{\rm R}(t)$, the spin current $\jsv \propto \partial_t \a_{\rm R}(t) \zv \times \mv$ gives rise to AC torques.
In the clean limit with $\hbar / \D \tau \ll 1$, the coefficient $\beta_{\rm R}$ is reduced to $\hbar / \D \tau$. Finally, the second term is exactly identical in form to the conventional current-induced nonadiabatic torque although $\tau$ corresponds to a different time scales, specifically, the relaxation time of the conduction electrons (spins) in the present (current-induced) case~\cite{Tatara08}. Assuming the above-mentioned material parameters for the metallic bilayer systems, we evaluate the values of $\js = (e/\hbar a) D_2$ and $\beta_{\rm R}$ as $\sim 2$~A/m and $\sim 0.07$, respectively. These values are large enough to induce the magnetization dynamics.

\subsection*{Application of oscillating DMI}
\begin{figure}[tb]
\centering
\includegraphics[scale=1.5, clip]{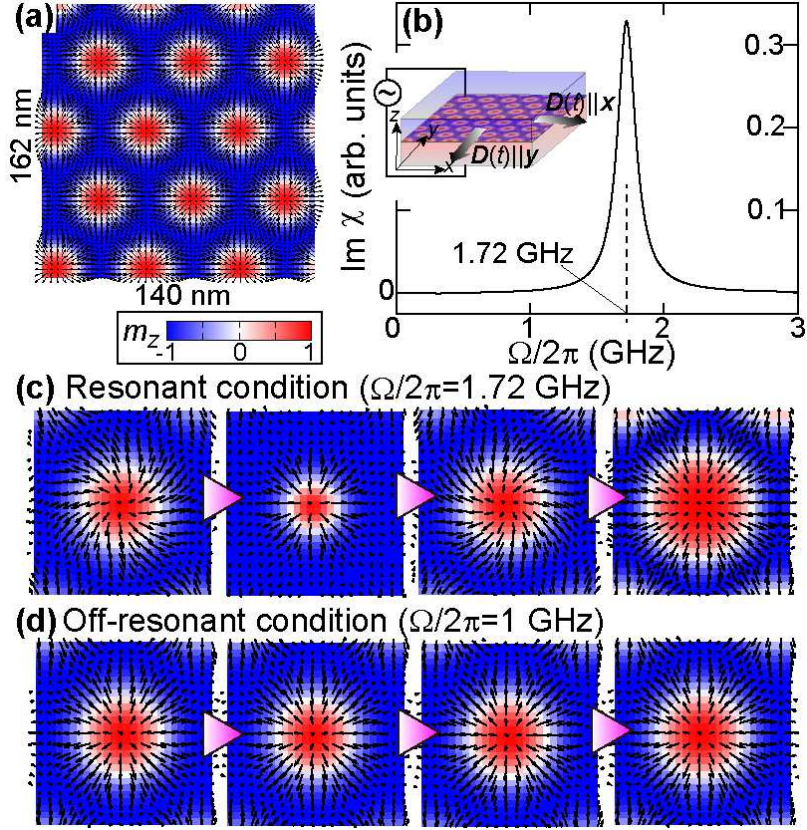}
\caption{Eigenmodes of N\'eel-type skyrmion crystal activated by an AC DMI. ({\bf a}) Skyrmion crystal with hexagonally packed N\'eel-type skyrmions. In-plane and out-of-plane components of the magnetizations are shown by arrows and color map, respectively. ({\bf b}) Imaginary part of the dynamical magnetoelectric susceptibility. ({\bf c}) Snapshots of the electrically activated breathing motion at $t=0$, $\pi / 2 \W$, $\pi / \W$ and $3 \pi / 2 \W$ for the resonant condition with $\W/2 \pi = 1.72$~GHz, and ({\bf d}) those for an off-resonant condition with $\W/2\pi = 1$~GHz.}
\label{fig2}
\end{figure}

The AC torques in the driven Rashba system may be exploited to activate resonances of the magnetic textures. For a demonstration, we numerically show that the breathing mode of a skyrmion crystal can be excited by the RSOI-mediated AC torques. We start with the continuum limit of the spin model for a two-dimensional magnet,
\begin{equation}
\H=\int{d^2r} \, \frac{J}{2}(\nabla \mv)^2
-\frac{\mu_{\rm B}\mu_0}{a^2} \int{d^2r} \, \mv \cdot {\bm H}.
\end{equation}
The model contains the ferromagnetic exchange interaction and the Zeeman interaction with an external magnetic field $\Hv = H \zv$. We adopt typical material parameters of $a=5$~\AA, $J=1$~meV and $\mu_0H=34$~mT. The dynamics of magnetizations obeys the Landau-Lifshitz-Gilbert equation
\begin{equation}
\dot{\mv}=-\gamma_{\rm m}
\mv \times \left(-\frac{a^2}{\gamma_{\rm m}\hbar} 
\frac{\delta \H}{\delta \mv} \right)
+\ag\mv\times \dot{\mv} + \Tv_1,
\end{equation}
where $\gamma_{\rm m}$ is the gyromagnetic ratio, and $\ag$ ($=0.04$) is the Gilbert-damping coefficient. We incorporate $\Tv_1$ as the Rashba-mediated torque whereas $\Tv_2$ and $\Tv_3$ are neglected because they are much smaller than $\Tv_1$ in the metallic bilayer systems. (Equation~\eqref{eq:torque} indicates that $\Tv_1$ ($\Tv_2$ and $\Tv_3$) originates from $D_1$ ($D_2$), and the ratio $D_2/D_1$ is 10$^{-12}$-10$^{-14}$ as mentioned above.) The coupling coefficient $D_1(t)$ is composed of a steady component $D_0$ and the time-dependent component $D_\ext(t)$ as $D_1(t)=D_0+D_\ext(t)$. We solve this equation numerically using the fourth-order Runge-Kutta method for a system of $140 \times 162$~nm$^2$, including $N=280 \times 324$ magnetizations, applying a periodic boundary condition. Without the AC electric voltage, the Rashba-mediated torque has a steady component only with $D_1=D_0$($=0.09$~meV). In this situation, the torque stabilizes the skyrmion crystal with hexagonally packed N\'eel-type skyrmions (Fig.~\ref{fig2}(a)) at low temperatures by effectively working as a static DMI, as exemplified by Eq.~(\ref{eqn:HDMI}).

Next, we simulated the dynamics of this skyrmion crystal in the presence of AC gate voltages by switching on the Rashba-mediated AC torque with $D_\ext(t) \ne 0$. The time evolution of the numerical simulation is performed for every $0.01$~ps. We first calculate the dynamical magnetoelectric susceptibility $\chi(\W)$ to identify the resonance frequency of this skyrmion crystal. We trace a time profile of the net magnetization $\Mv(t) = (1 / Na^2) \int{d^2r} \, \mv(\rv, t)$ and $\Delta \Mv(t) = \Mv(t) -\Mv(0)$ after applying a short electric-field pulse by switching on the time-dependent component of the torque $D_\ext(t) = 0.05 D_0$ for an interval of $0 \leq t \leq 1$. We obtain its spectrum (Fig.~\ref{fig2}(b)) by calculating the Fourier transform of $\Delta \Mv(t)$. From this spectrum, we find that the skyrmion crystal has a resonance at $\W / 2\pi = 1.72$~GHz. Note that the resonance frequency (the time period) of the skyrmion eigenmode is determined by the spin-wave gap which is proportional to $D_0^2$ ($D_0^{-2}$).

We then apply a microwave electric field to this skyrmion-hosting magnetic bilayer system by switching on the AC component of the torque $D_\ext(t)=0.05 D_0 \sin{(\W t)}$. We examined both resonance with $\W/2\pi=1.72$~GHz and off-resonance with $\W/2\pi=1$~GHz.
For the former, we observe a breathing-type mode where all the skyrmions constituting the skyrmion crystal uniformly contracted and expanded periodically (Fig.~\ref{fig2}(c)).
For the latter, the induced breathing oscillation of the skyrmions is small.
We find that the periodically modulated skyrmion diameter is nearly proportional to the time-dependent DMI coefficient. The present simulation demonstrated the electrical activation of a skyrmion resonance. The induced dynamical DMI is also expected to realize the electrical creation of skyrmions~\cite{Mochizuki15,Mochizuki16,Schott17,HuangP18,Kurchkov18,WangL18}. The electrical writing of skyrmions with the RSOI-mediated DMI in the field polarized ferromagnetic state and even the helical state should be established in the future study.

\section*{Discussion}
In summary, we have theoretically derived a precise formula of the spin torque in a time-varying Rashba electron system driven by the AC gate voltage.
The obtained formula contains not only the spin-transfer torque but also the $\beta$-term torque associated with the AC spin current that is proportional to the time derivative of the RSOI.
These AC torques can excite resonance of magnetic textures through acting as an interfacial AC DMI.
Indeed, we have numerically demonstrated that the effective AC DMI can activate the breathing mode of magnetic skyrmions.
We have confirmed that not only the crystallized magnetic skyrmions (skyrmion crystal) but also isolated skyrmions in ferromagnets can be excited resonantly by application of a microwave field, which offers a better experimental feasibility because a lot of ferromagnet/heavy-metal bilayer systems turned out to host magnetic skyrmions as topological defects~\cite{Fert13,SWoo16,GYu17,WJiang15}.
Recent theoretical studies revealed that activation of the skyrmion breathing mode under application of a magnetic field inclined from the vertical direction induces translational motion of the skyrmions~\cite{Wang15, Takeuchi18,Ikka18}, which provides a means to drive magnetic skyrmions electrically with a low energy consumption.
Our finding provides a promising technique to manipulate noncollinear magnetic textures with a great efficiency that have potential applications in memory, logic, and microwave devices.

There are several types of devices to realize the proposed effects. The devices must have two important features, i.e., (1) an interface that hosts the SOI due to the broken inversion symmetry and (2) insulating nature to prevent the electric-current flow. One possible type of device is a ferromagnet/insulator bilayer system~\cite{Nawaoka15,Srivastava18}. On the contrary, we proposed another type of device with ferromagnet/heavy-metal bilayer fabricated on an insulating substrate. In the latter system, we expect much stronger SOI because of the heavy-metal layer. In this case, the required insulating nature is taken up by the insulating substrate. The issue which system is appropriate is left for future study. The RSOI is originally very strong in the latter system, but its electric tunability may be low because the applied electric field mainly acts on the heavy-metal/insulator layer but not on the ferromagnet/heavy-metal interface. In addition, due to the short screening length in metal, the electric field decays quickly and may hardly reach the interface that hosts the RSOI. On the other hand, the ferromagnet/insulator bilayer system can originally have a weak RSOI only, but its electric tunability can be large because the applied electric field directly acts on the ferromagnet/insulator interface that hosts the RSOI.

It should be also mentioned that several types of bilayer systems with interfacial DMI have been intensively studied recently, where driven spin torques and generations of skyrmion-type noncollinear magnetic textures using the interfacial DMI have been experimentally demonstrated not only for ferromagnet/heavy-metal systems but also for ferromagnet/transition-metal-dichalcogenide systems~\cite{RHLiu14} and ferromagnet/topological-insulator systems~\cite{WLv18}.
In the latter two cases, the SOI is much more complicated than the simple Rashba model for the ferromagnet/heavy-metal system considered in the present study~\cite{Belashchenko18}, and it is unclear which of many terms emerged at the interfaces of these systems can be modulated by AC gate voltage in reality~\cite{Gmitra16}.
These problems require further investigations, which are left for future researches. In experiments, we need to take care of magnetic anisotropies in the ferromagnetic layer because the stability and the resonance modes of skyrmions are sensitively affected by them~\cite{Nakamura18}. Our results will provide a firm basis and a good starting point for future experimental and theoretical studies.

\section*{Methods}
\subsection*{Diagonalization of exchange interaction}
We need to calculate the electron spin density vector $\sv$ to obtain the spin torque.
However, as the exchange coupling $\D$ is generally large compared with other energy scales such as the kinetic energy and the RSOI, we cannot regard it as a perturbation.
Instead, we need to perform a diagonalization of the exchange interaction term~\cite{Tatara08,Ho15}.
For this purpose, we employ a $2\times2$ unitary matrix $\U \equiv \nv \cdot \pauliv$ with $\nv = (\sin{(\theta/2)} \cos{\phi}, \sin{(\theta/2)} \sin{\phi}, \cos{(\theta/2)})$.
Here, $\theta$ and $\phi$ are the polar and azimuthal angles of the local magnetization vector $\mv = (\sin{\theta} \cos{\phi}, \sin{\theta} \sin{\phi}, \cos{\theta})$.
As relation $\mv \cdot \U^\dagger \pauliv \U = \pauli^z$ holds, the exchange interaction term may be diagonalized using a new operator of the conduction electron $\Psi$ expressed in the local coordinates rotating along with the spatially modulated magnetization vectors.
Hereafter, this coordinate system is referred to as the rotated spin frame.
The new operator $\Psi$ is related to the original operator $\psi$ defined in the global coordinates via a unitary transformation as $\psi = \U \Psi$.
Finally, the total Hamiltonian is rewritten with the $\Psi$ operators as
\begin{eqnarray*}
H
&=&
\frac{1}{2 \me} \int{d^2r} \, \big| \big[ \pv +e \Av^\alpha(\rv,t) \pauli^\alpha \big] \Psi(\rv,t) \big|^2
-\Ef \int{d^2r} \, \Psi^\dagger(\rv,t) \Psi(\rv,t)
\\
&&
+\D \int{d^2r} \, \Psi^\dagger(\rv,t) \pauli^z \Psi(\rv,t)
\\
&&
-\frac{\me \a_{\rm R}^2(t)}{\hbar^2} \int{d^2r} \, \Psi^\dagger(\rv,t) \Psi(\rv,t)
+\int{d^2r} \, v_{\rm imp}(\rv) \Psi^\dagger(\rv,t) \Psi(\rv,t),
\end{eqnarray*}
where $-e$ ($<0$) is the electron charge.
The non-Abelian gauge potential $\A$ appears as a by-product of the diagonalization of the exchange interaction.
This gauge potential contains two contributions~\cite{Ho15}, denoted $\A = \A^\ex +\A^\so$.
The first term $\A^\ex$ is a gauge potential originating from the spatial variation of the magnetization structure whereas the second term $\A^\so$ comes from the RSOI.
Each gauge potential is defined as
\begin{align*}
\A_\mu^{\ex \alpha} \pauli^\alpha
&\equiv
-\frac{i \hbar}{e} \U^\dagger \nabla_\mu \U
=
\frac{\hbar}{e} (\nv \times \nabla_\mu \nv)^\alpha \pauli^\alpha,
\\
\A_\mu^{\so \alpha} \pauli^\alpha
&\equiv
-\frac{\me \a_{\rm R}}{e \hbar} \epsilon_{\mu \beta z} \U^\dagger 
\pauli^\beta \U
=
-\frac{\me \a_{\rm R}}{e \hbar} \epsilon_{\mu \beta z} \R^{\alpha\beta} \pauli^\alpha,
\end{align*}
with $\R^{\alpha\beta} \pauli^\alpha \equiv \U^\dagger \pauli^\beta \U = (2 n^\alpha n^\beta -\delta_{\alpha\beta}) \pauli^\alpha$.
Here $\R^{\alpha\beta}$ is an element of a $3 \times 3$ orthogonal matrix.
The symbols $\delta_{\alpha\beta}$ and $\epsilon_{\alpha\beta\gamma}$ denote the Kronecker delta and the Levi-Civita antisymmetric tensor, respectively.
In the rotated spin frame, the $\Psi$-electron spin density is given by $S^\alpha = \langle{\Psi^\dagger \pauli^\alpha \Psi}\rangle$.
The spin density $S$ in the rotated spin frame is related to the spin density $s$ in the original frame via $\R$ as $s^\alpha = \R^{\alpha\beta} S^\beta$.
Therefore, the spin torque $\T$ is rewritten using $S^\pm = S^x \pm i S^y$,
\begin{equation*}
\T^\alpha
=
\frac{\D a^2}{2 \hbar} \epsilon_{\alpha\beta\gamma} m^\beta
\sum_{\sigma = \pm} (\R^{\gamma x} -i \sigma \R^{\gamma y}) S^\sigma.
\end{equation*}

In the calculation, we consider the impurity potential given by
\begin{equation*}
v_{\rm imp}(\rv) = u_{\rm imp} \sum_i \delta(\rv -{\bm R}_i),
\end{equation*}
where $u_{\rm imp}$ denotes the strength of impurity scattering,
${\bm R}_i$ the positions of the impurities,
and $\delta(\rv)$ the Dirac delta function.
When we take an average over the impurity positions as 
\begin{align*}
& \overline{v_{\rm imp}(\rv)} = 0,
\\
& \overline{v_{\rm imp}(\rv) v_{\rm imp}(\rv')} = n_{\rm imp} u_{\rm imp}^2 \delta(\rv-\rv')
\end{align*}
the relaxation time of electrons is given by $\tau = \hbar / 2 \pi \nu_{\rm e} n_{\rm imp} u_{\rm imp}^2$ in the first Born approximation.
Here, $n_{\rm imp}$ denotes the concentration of impurities and $\nu_{\rm e}=\me/2\pi \hbar^2$ the density of state.

\subsection*{Calculation of spin torque arising from AC RSOI}
The $\Psi$-electron spin density $S^\pm$ is written in terms of the path-ordered Green function
\begin{equation*}
S^\pm(\rv,t)
=
-i \hbar \tr
\Big[ \pauli^\pm \hat{G}^<(\rv,t; \rv,t) \Big],
\end{equation*}
where $\pauli^\pm = \pauli^x \pm i \pauli^y$ and $\tr$ signifies the trace over the spin indices.
The lesser component of the path-ordered Green function~\cite{HaugJauho} is represented by 
$\hat{G}^<(\rv,t; \rv',t') = (i/\hbar) \langle{\Psi^\dag(\rv',t') \Psi(\rv,t)}\rangle$.
In the present system, the Dyson equation is given as
\begin{equation*}
\hat{G}(\rv,t; \rv',t')
=
\hat{g}(\rv-\rv',t-t')
+\int_C{dt''}\, \int{d^2r''}\,
\hat{g}(\rv-\rv'',t-t'') \hat{V}(\rv'',t'') \hat{G}(\rv'',t'';\rv',t'),
\end{equation*}
where $C$ denotes the Keldysh contour and $\hat{V}$ is defined as
\begin{eqnarray*}
\hat{V}(\rv'',t'')
&=&
-\frac{i e \hbar}{2 \me}
\bigg[ \frac{\del}{\del r_\mu''} \A_\mu^\alpha(\rv'',t'') +\A_\mu^\alpha(\rv'',t'') \frac{\del}{\del r_\mu''} \bigg] \pauli^\alpha
\\
&&
+\frac{e^2}{2 \me} \A_\mu^\alpha(\rv'',t'') \A_\mu^\alpha(\rv'',t'') \hat{\rm I}
-\frac{\me}{\hbar^2} \a_{\rm R}^2(t'') \hat{\rm I}
+v_{\rm imp}(\rv'') \hat{\rm I}.
\end{eqnarray*}
Here $\g$ denotes the noninteracting Green function given in the Fourier space as $\g_{\kv,\w} = (1/2) \sum_{\sigma = \pm} (\hat{\rm I} +\sigma \pauli^z) g_{\kv,\w,\sigma}$ where $\hat{\rm I}$ is the identity matrix.
The superscript symbol $<$ represents the relation
$[\int_C{dt''}\, \hat{G}_1(t,t'') \hat{G}_2(t'',t') ]^< = \int_{-\infty}^\infty{dt''}\, [\hat{G}^{\rm r}_1(t,t'') \hat{G}^<_2(t'',t') +\hat{G}^<_1(t,t'') \hat{G}^{\rm a}_2(t'',t')]$~\cite{HaugJauho}.
The retarded, advanced, and lesser Green functions ($\gr$, $\ga$ and $g^<$) are mutually related by $g_{\kv,\w,\sigma}^< = f_\w (\ga_{\kv,\w,\sigma} -\gr_{\kv,\w,\sigma})$ where $f_\w$ is the Fermi distribution function.
The retarded (advanced) Green function is defined as $\gr_{\kv,\w,\sigma} = (\ga_{\kv,\w,\sigma})^* = 1 / (\hbar \w -\Ek +\varepsilon_{{\rm F} \sigma} +i \eta)$ where $\Ek = \hbar^2 \kv^2 / 2 \me$ and $\varepsilon_{{\rm F} \sigma} = \Ef -\sigma \D$.
The spin density $S^\pm$ can be obtained by iteration of this equation.
The dominant contributions are given by the first-order perturbation expansions in $\A^\so$ and up to first order in $\A^\ex$.
After some algebra, this equation is reduced to (see Supplementary Materials for details)
\begin{equation*}
S^\pm
=
-\frac{2 e}{\hbar \D a} \Big[ D_1 -(1 \mp i \beta_{\rm R}) D_2 \Big]
\Big( \mv \times \Av^{\ex \pm} \Big)_z,
\end{equation*}
where $\A^{\ex \pm} = \A^{\ex x} \pm i \A^{\ex y}$.
Substituting this result into the definition of the spin torque and using the relations
$\sum_{\sigma = \pm} (\R^{\alpha x} -i \sigma \R^{\alpha y}) \A_\mu^{\ex \sigma}
=
-(\hbar / e) (\mv \times \nabla_\mu \mv)^\alpha$
and
$\sum_{\sigma = \pm} (\R^{\alpha x} -i \sigma \R^{\alpha y}) i \sigma \A_\mu^{\ex \sigma}
=
(\hbar / e) \nabla_\mu m^\alpha$,
we thus have obtained the result given in Eqs.~(\ref{eq:torque}-\ref{eq:beta}).

\subsection*{Note on the Numerical Simulations}
Our numerical simulation with the LLG equation corresponds to the micromagnetic simulation based on the continuum spin model with the exchange stiffness $A=4.0\times10^{-14}$ [Jm$^{-1}$], the continuum DM parameter $B=1.44\times 10^{-5}$ [Jm$^{-2}$], the magnetic field $\mu_0H_z$=34 mT, and the saturation magnetization $M_{\rm s}=3.5\times 10^4$ [Am$^{-1}$] for a system size of 140 nm $\times$ 162 nm $\times$ 2 nm. This simulation can be performed with commercial or free softwares such as OOMMF and mumax. In the present study, the continuum spin model is mapped to the lattice spin model by dividing the continuum space into identical rectangular cells. More concretely, dividing the continuum space into identical rectangular cells of 0.5 nm$\times$0.5 nm$\times$2 nm, we obtain the lattice spin model with the normalized magnetization vectors $\bm m_i$ where the exchange coupling $J$=1 meV, the DM parameter $D_0/J$=0.09, and the magnetic field $\mu_{\rm B}\mu_0H_z/J$=0.004.

\section*{Acknowledgements}
This work was supported by JSPS KAKENHI (Grant No. 17H02924 and No. 16H06345), Waseda University Grant for Special Research Projects (Project Nos. 2017S-101, 2018K-257), and JST PRESTO (Grant No. JPMJPR132A).

\newpage

\noindent
{\fontsize{20}{12}\usefont{OT1}{phv}{b}{n}
Supplementary Material for
``Electrically driven spin torque and dynamical Dzyaloshinskii-Moriya \\ interaction in magnetic bilayer systems''}

\subsection*{Calculation of $\Psi$-electron spin density $S^\pm$}
\begin{figure}[ht]
\centering
\includegraphics[scale=1.2, clip]{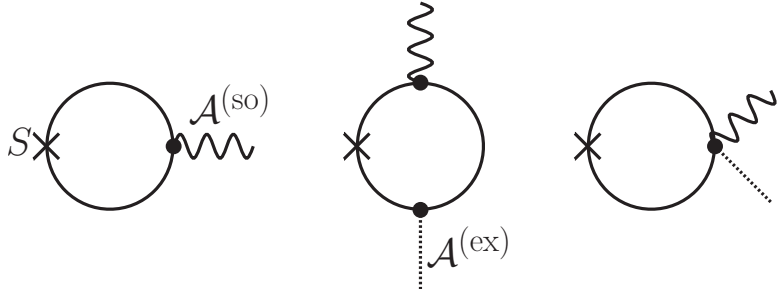}
\caption{
Diagrammatic representations of the dominant contributions of the $\Psi$-electron spin density $S$.
The solid, wavy, and dotted lines represent the Green function, the RSOI $\A^\so$, and the exchange interaction $\A^\ex$ in the rotated spin frame, respectively.
}
\label{smfig1}
\end{figure}

We provide details of the derivation of the $\Psi$-electron spin density $S^\pm$.
In Fig.~\ref{smfig1}, we present the Feynman diagrams associated with the $\Psi$-electron spin density induced by the time-dependent RSOI.
Here the vertex corrections due to the nonmagnetic impurity scatterings are not considered because they are negligible in the present case.

Expanding the lesser component with respect to $\qv$ and $\Omega$, the spin density $S^\pm (\bm q,\Omega)$ is written
\begin{eqnarray}
S^\pm(\qv,\W)
&=&
\frac{i e \hbar^4}{\me^2} \bigg\{
\pm q_\nu \A_\mu^{\so \pm}(\qv,\W)
-\frac{2 e}{\hbar} \sum_{\qv'} \Big[
\A_\mu^{\so \pm}(\qv',\W) \A_\nu^{\ex z}(\qv-\qv')
+\A_\mu^{\so z}(\qv',\W) \A_\nu^{\ex \pm}(\qv-\qv')
\Big]
\bigg\}
\notag
\\
&&
\times
\sum_{\sigma = \pm} \sigma \sum_\kv \sum_\w f_\w k_\mu k_\nu
\Big[
\ga_{\kv, \w, -\sigma} (\ga_{\kv, \w, \sigma})^2 -\gr_{\kv, \w, -\sigma} (\gr_{\kv, \w, \sigma})^2
\Big]
\notag
\\
&&
+\frac{i e \hbar^4}{2 \me^2} \W \bigg[
\pm q_\nu \A_\mu^{\so \pm}(\qv,\W)
-\frac{2 e}{\hbar} \sum_{\qv'} \A_\mu^{\so \pm}(\qv',\W) \A_\nu^{\ex z}(\qv-\qv')
\bigg]
\notag
\\
&&
\times
\sum_{\sigma = \pm} (\sigma \mp 1) \sum_\kv \sum_\w k_\mu k_\nu \bigg\{
\frac{d f_\w}{d \w} \bigg[
\gr_{\kv, \w, \sigma} (\ga_{\kv, \w, -\sigma})^2 -(\gr_{\kv, \w, \sigma})^2 \ga_{\kv, \w, -\sigma}
\notag
\\
&&
+\frac{1}{2} \ga_{\kv,\w,\sigma} (\ga_{\kv,\w,\sigma} -\ga_{\kv,\w,-\sigma}) \ga_{\kv,\w,-\sigma}
+\frac{1}{2} \gr_{\kv,\w,\sigma} (\gr_{\kv,\w,\sigma} -\gr_{\kv,\w,-\sigma}) \gr_{\kv,\w,-\sigma}
\bigg]
\notag
\\
&&
-\hbar f_\w \Big[
\ga_{\kv,\w,-\sigma} \ga_{\kv,\w,\sigma} (\ga_{\kv,\w,\sigma} -\ga_{\kv,\w,-\sigma})^2
+(\ga_{\kv,\w,-\sigma})^2 (\ga_{\kv,\w,\sigma})^2
\notag
\\
&&
-\gr_{\kv,\w,-\sigma} \gr_{\kv,\w,\sigma} (\gr_{\kv,\w,\sigma} -\gr_{\kv,\w,-\sigma})^2
-(\gr_{\kv,\w,-\sigma})^2 (\gr_{\kv,\w,\sigma})^2
\Big]
\bigg\}
\notag
\\
&&
+\frac{i e^2 \hbar^3}{\me^2} \W \sum_{\qv'} \A_\mu^{\so z}(\qv',\W) \A_\nu^{\ex \pm}(\qv-\qv')
\notag
\\
&&
\times
\sum_{\sigma = \pm} (\sigma \mp 1) \sum_\kv \sum_\w k_\mu k_\nu \bigg\{
\frac{d f_\w}{d \w} \bigg[
\gr_{\kv, \w, \sigma} \ga_{\kv, \w, -\sigma} \ga_{\kv, \w, \sigma}
-\gr_{\kv, \w, -\sigma} \gr_{\kv, \w, \sigma} \ga_{\kv, \w, -\sigma}
\notag
\\
&&
-\frac{1}{2} \ga_{\kv,\w,\sigma} (\ga_{\kv,\w,\sigma} -\ga_{\kv,\w,-\sigma}) \ga_{\kv,\w,-\sigma}
-\frac{1}{2} \gr_{\kv,\w,\sigma} (\gr_{\kv,\w,\sigma} -\gr_{\kv,\w,-\sigma}) \gr_{\kv,\w,-\sigma}
\bigg]
\notag
\\
&&
-\hbar f_\w \Big[
(\ga_{\kv,\w,-\sigma})^2 (\ga_{\kv,\w,\sigma})^2
-(\gr_{\kv,\w,-\sigma})^2 (\gr_{\kv,\w,\sigma})^2
\Big]
\bigg\}.
\end{eqnarray}
Using the following relation,
\begin{equation}
\pm i \nabla_\nu \A_\mu^{\so \pm}
+\frac{2 e}{\hbar} \A_\mu^{\so \pm} \A_\nu^{\ex z}
=
\frac{2 e}{\hbar} \A_\mu^{\so z} \A_\nu^{\ex \pm},
\end{equation}
the spin density $S^\pm (\rv, t)$ is rewritten as
\begin{eqnarray}
S^\pm(\rv,t)
&=&
\frac{8 e^2 \hbar^3}{\me^2} \A_\mu^{\so z}(\rv,t) \A_\nu^{\ex \pm}(\rv)
\Im \sum_{\sigma = \pm} \sigma \sum_\kv \sum_\w f_\w k_\mu k_\nu
\ga_{\kv, \w, -\sigma} (\ga_{\kv, \w, \sigma})^2
\notag
\\
&&
+\frac{4 e^2 \hbar^3 \D}{\me^2} \frac{\del \A_\mu^{\so z}(\rv,t)}{\del t} \A_\nu^{\ex \pm}(\rv)
\sum_{\sigma = \pm} \sum_\kv \sum_\w k_\mu k_\nu \bigg\{
\frac{d f_\w}{d \w}
\Re (\ga_{\kv,\w,\sigma} -\gr_{\kv, \w, \sigma}) (\ga_{\kv, \w, -\sigma})^2 \ga_{\kv, \w, \sigma}
\notag
\\
&&
\pm i \Im \bigg[
\sigma \frac{d f_\w}{d \w} \gr_{\kv, \w, \sigma} (\ga_{\kv, \w, -\sigma})^2 \ga_{\kv, \w, \sigma}
+2 \hbar \D f_\w (\ga_{\kv,\w,-\sigma})^3 (\ga_{\kv,\w,\sigma})^3
\bigg]
\bigg\}.
\end{eqnarray}
Here we neglect contributions from the products of $\ga$ to calculate terms proportional to the first order in $\W$ as they only give small corrections compared with those from the products of $\gr$ and $\ga$.
Therefore, the spin density reduces to
\begin{eqnarray}
S^\pm(\rv,t)
&=&
\frac{8 e^2 \hbar^3}{\me^2} \A_\nu^{\ex \pm}(\rv) \bigg[
\A_\mu^{\so z}(\rv,t)
\Im \sum_{\sigma = \pm} \sigma \sum_\kv \sum_\w f_\w k_\mu k_\nu
\ga_{\kv, \w, -\sigma} (\ga_{\kv, \w, \sigma})^2
\notag
\\
&&
-\frac{\D}{2} \frac{\del \A_\mu^{\so z}(\rv,t)}{\del t}
\sum_{\sigma = \pm} (\Re \mp i \sigma \Im) \sum_\kv \sum_\w \frac{d f_\w}{d \w} k_\mu k_\nu
\gr_{\kv, \w, \sigma} (\ga_{\kv, \w, -\sigma})^2 \ga_{\kv, \w, \sigma}
\bigg]
\notag
\\
&=&
\frac{8 e^2 \hbar}{\me} \A_\nu^{\ex \pm}(\rv) \bigg\{
C_{\mu\nu}^{(1)} \A_\mu^{\so z}(\rv,t)
-\frac{\D}{2} \Big[ C_{\mu\nu}^{(2)} \mp i C_{\mu\nu}^{(3)} \Big] \frac{\del \A_\mu^{\so z}(\rv,t)}{\del t}
\bigg\}.
\end{eqnarray}
The summations over $\kv$ and $\w$ are performed in the following manner,
\begin{eqnarray}
C_{\mu\nu}^{(1)}
&=&
\frac{\hbar^2}{\me}
\Im \sum_{\sigma = \pm} \sigma \sum_\kv \sum_\w f_\w k_\mu k_\nu
\ga_{\kv, \w, -\sigma} (\ga_{\kv, \w, \sigma})^2
\notag
\\
&=&
\frac{\nu_{\rm e}}{2 \pi} \delta_{\mu \nu}
\Im \sum_{\sigma = \pm} \int_0^\infty{d\E}\, \int_{-\infty}^0{d\w}\,
\frac{\E}{(\hbar \w -\E +\E_{{\rm F}, -\sigma} -i \eta) (\hbar \w -\E +\E_{{\rm F}, \sigma} -i \eta)^2}
\notag
\\
&=&
\frac{\nu_{\rm e}}{4 \pi \hbar \D^2} \delta_{\mu \nu}
\Im \sum_{\sigma = \pm} \sigma \int_{-\infty}^0{d\E}\, \int_{-\infty}^0{d\w}\,
\frac{\E -\sigma \D}{\w +\E +\E_{{\rm F}, \sigma} -i \eta}
\notag
\\
&=&
\frac{\nu_{\rm e}}{4 \pi \hbar \D^2} \delta_{\mu \nu}
\Im \sum_{\sigma = \pm} \sigma \int_{-\infty}^{-\sigma \D}{d\E}\, \int_{-\infty}^0{d\w}\,
\frac{\E}{\w +\E +\Ef -i \eta}
\notag
\\
&=&
-\frac{\nu_{\rm e}}{4 \pi \hbar \D^2} \delta_{\mu \nu}
\Im \int_{-\D}^{\D}{d\E}\, \int_{-\infty}^0{d\w}\,
\frac{\E}{\w +\E +\Ef -i \eta}
\notag
\\
&=&
-\frac{\nu_{\rm e}}{4 \pi \hbar \D^2} \delta_{\mu \nu}
\Im \int_{-\D}^{\D}{d\E}\,
\E \Big[ \ln{(\E +\Ef -i \eta)} +i \pi \Big]
\notag
\\
&=&
-\frac{\nu_{\rm e} \eta}{16 \pi \tau \D} \delta_{\mu \nu}
\bigg\{
\frac{\Ef}{\D} \ln{\bigg[\frac{(\Ef +\D)^2 +\eta^2}{(\Ef -\D)^2 +\eta^2}\bigg]}
-2
\notag
\\
&&
+\frac{\Ef^2 -\D^2 -\eta^2}{\D \eta} \bigg[
\tan^{-1}{\bigg(\frac{\eta}{\Ef +\D}\bigg)}
-\tan^{-1}{\bigg(\frac{\eta}{\Ef -\D}\bigg)}
\bigg]
\bigg\},
\end{eqnarray}
\begin{eqnarray}
C_{\mu \nu}^{(2)}
&=&
\frac{\hbar^2}{\me}
\Re \sum_{\sigma = \pm}
\sum_\kv \sum_\w \frac{d f_\w}{d \w} k_\mu k_\nu
\gr_{\kv, \w, \sigma} (\ga_{\kv, \w, -\sigma})^2 \ga_{\kv, \w, \sigma}
\notag
\\
&=&
\frac{\nu_{\rm e}}{2 \pi} \delta_{\mu\nu}
\Re \sum_{\sigma = \pm}
\int_{-\infty}^0{d\E} \,
\frac{\E}{(\E +\E_{{\rm F}, \sigma} +i \eta) (\E +\E_{{\rm F}, -\sigma} -i \eta)^2 (\E +\E_{{\rm F}, \sigma} -i \eta)}
\notag
\\
&=&
\frac{\nu_{\rm e}}{8 \pi \D} \delta_{\mu\nu}
\Re \sum_{\sigma = \pm} \sigma \int_{-\infty}^0{d\E} \, \bigg[
\frac{\sigma \Ef \D}{i 2 \eta (\sigma \D -i \eta)^2}
\bigg(
\frac{1}{\E +\Ef -\sigma \D +i \eta}
-\frac{1}{\E +\Ef +\sigma \D -i \eta}
\bigg)
\notag
\\
&&
-\frac{1}{i 2 \eta}
\bigg(
\frac{1}{\E +\Ef -\sigma \D +i \eta}
-\frac{1}{\E +\Ef -\sigma \D -i \eta}
\bigg)
-\frac{\Ef -\sigma \D +i \eta}{\sigma \D -i \eta}
\frac{1}{(\E +\Ef -\sigma \D +i \eta)^2}
\bigg]
\notag
\\
&=&
\frac{\nu_{\rm e}}{8 \pi \D} \delta_{\mu\nu}
\Re \sum_{\sigma = \pm} \sigma \bigg\{
\frac{\sigma \Ef \D}{i 2 \eta (\sigma \D -i \eta)^2}
\bigg[
\ln{\bigg(\frac{\Ef -\sigma \D +i \eta}{\Ef +\sigma \D -i \eta}\bigg)}
-i 2 \pi
\bigg]
\notag
\\
&&
-\frac{1}{i 2 \eta}
\bigg[
\ln{\bigg(\frac{\Ef -\sigma \D +i \eta}{\Ef -\sigma \D -i \eta}\bigg)}
-i 2 \pi
\bigg]
+\frac{1}{\sigma \D -i \eta}
\bigg\}
\notag
\\
&=&
-\frac{\nu_{\rm e} \tau \Ef (\D^2 -\eta^2)}{2 \hbar (\D^2 +\eta^2)^2}
\delta_{\mu\nu} \bigg\{
1
-\frac{\eta (\D^2 +\eta^2)}{\pi \Ef (\D^2 -\eta^2)}
+\frac{\eta \D}{2 \pi (\D^2 -\eta^2)}
\ln{\bigg[\frac{(\Ef +\D)^2 +\eta^2}{(\Ef -\D)^2 +\eta^2}\bigg]}
\notag
\\
&&
-\frac{1}{2 \pi}
\bigg[
\tan^{-1}{\bigg(\frac{\eta}{\Ef +\D}\bigg)}
+\tan^{-1}{\bigg(\frac{\eta}{\Ef -\D}\bigg)}
\bigg]
\notag
\\
&&
-\frac{(\D^2 +\eta^2)^2}{2 \pi \Ef \D (\D^2 -\eta^2)}
\bigg[
\tan^{-1}{\bigg(\frac{\eta}{\Ef +\D}\bigg)}
-\tan^{-1}{\bigg(\frac{\eta}{\Ef -\D}\bigg)}
\bigg]
\bigg\},
\end{eqnarray}
\begin{eqnarray}
C_{\mu\nu}^{(3)}
&=&
\frac{\hbar^2}{\me}
\Im \sum_{\sigma = \pm} \sigma
\sum_\kv \sum_\w \frac{d f_\w}{d \w} k_\mu k_\nu
\gr_{\kv, \w, \sigma} (\ga_{\kv, \w, -\sigma})^2 \ga_{\kv, \w, \sigma}
\notag
\\
&=&
\frac{\nu_{\rm e}}{8 \pi \D} \delta_{\mu\nu}
\Im \sum_{\sigma = \pm} \int_{-\infty}^0{d\E} \, \bigg[
\frac{\Ef (\sigma 2 \D -i \eta)}{\sigma 2 \D (\sigma \D -i \eta)^2}
\bigg(
\frac{1}{\E +\Ef -\sigma \D +i \eta}
-\frac{1}{\E +\Ef +\sigma \D -i \eta}
\bigg)
\notag
\\
&&
-\frac{1}{\sigma 2 \D}
\bigg(
\frac{1}{\E +\Ef -\sigma \D +i \eta}
-\frac{1}{\E +\Ef -\sigma \D -i \eta}
\bigg)
-\frac{\Ef -\sigma \D +i \eta}{\sigma \D -i \eta}
\frac{1}{(\E +\Ef -\sigma \D +i \eta)^2}
\bigg]
\notag
\\
&=&
\frac{\nu_{\rm e}}{8 \pi \D} \delta_{\mu\nu}
\Im \sum_{\sigma = \pm} \bigg\{
\frac{\Ef (\sigma 2 \D -i \eta)}{\sigma 2 \D (\sigma \D -i \eta)^2}
\bigg[
\ln{\bigg(\frac{\Ef -\sigma \D +i \eta}{\Ef +\sigma \D -i \eta}\bigg)}
-i 2 \pi
\bigg]
\notag
\\
&&
-\frac{1}{\sigma 2 \D}
\bigg[
\ln{\bigg(\frac{\Ef -\sigma \D +i \eta}{\Ef -\sigma \D -i \eta}\bigg)}
-i 2 \pi
\bigg]
+\frac{1}{\sigma \D -i \eta}
\bigg\}
\notag
\\
&=&
-\frac{\nu_{\rm e} \Ef \D}{2 (\D^2 +\eta^2)^2}
\delta_{\mu\nu} \bigg\{
1
-\frac{\eta (\D^2 +\eta^2)}{2 \pi \Ef \D^2}
+\frac{\eta (3 \D^2 +\eta^2)}{8 \pi \D^3}
\ln{\bigg[\frac{(\Ef +\D)^2 +\eta^2}{(\Ef -\D)^2 +\eta^2}\bigg]}
\notag
\\
&&
-\frac{1}{2 \pi}
\bigg[
\tan^{-1}{\bigg(\frac{\eta}{\Ef +\D}\bigg)}
+\tan^{-1}{\bigg(\frac{\eta}{\Ef -\D}\bigg)}
\bigg]
\notag
\\
&&
-\frac{(\D^2 +\eta^2)^2}{4 \pi \Ef \D^3}
\bigg[
\tan^{-1}{\bigg(\frac{\eta}{\Ef +\D}\bigg)}
-\tan^{-1}{\bigg(\frac{\eta}{\Ef -\D}\bigg)}
\bigg]
\bigg\}.
\end{eqnarray}

\end{document}